# On the U(2) Lattice Gauge Theory

Claude Roiesnel*

*Centre de Physique Théorique,*
*Centre National de la Recherche Scientifique, UPR A0014,*
*Ecole Polytechnique, 91128 Palaiseau Cedex, FRANCE*

September 1995

hep-lat/9509092   28 Sep 1995

## Abstract

We study the U(2) lattice gauge theory in the pure gauge sector using the simplest action, with determinant and fundamental terms, having the naive continuum limit of SU(2)×U(1). We determine part of the phase diagram of the model and find a first-order critical line which goes through the U(1) critical point. We show how to deduce both the order parameter of the first-order transition and the U(2) renormalization group flow from the lattice potential in the determinant and fundamental representations. We give evidence that a Monte-Carlo simulation of the model is indeed consistent with the symmetric SU(2)×U(1) continuum limit in the weak coupling pertubative regime.

## 1  Introduction

A description of the weak and electromagnetic interactions by a local gauge theory based on the group U(2) is consistent with the quantum numbers of the observed elementary particles. Indeed the global structure of the standard model can be inferred from the observed patterns of the irreducible representations of matter fields which obey the quantization rule:

$$Q = T_3 + \frac{Y}{2} \qquad (1)$$

where $T$ and $Y$ are the weak SU(2) isospin and weak U(1) hypercharge respectively and $Q$ is the electric charge. It has long been emphasized [1] that the global structure of groups is relevant to particle physics.

On the lattice, the formulation of the U(2) gauge theory differs from that of the SU(2)×U(1) gauge theory already in the pure gauge sector. Indeed U(2) is isomorphic to the quotient group

$$\mathrm{U}(2) = \frac{\mathrm{SU}(2) \times \mathrm{U}(1)}{\mathrm{Z}(2)} \qquad (2)$$

which means that each element $U$ of the group U(2) can be represented as a pair $(e^{i\phi}, V) \in$ U(1)×SU(2), with the identification $(e^{i\phi}, V) \equiv (-e^{i\phi}, -V)$. Wilson's lattice gauge invariant formulation [2] depends explicitly on the topological structure of the gauge group because the degrees of freedom belong to the group manifold, not its Lie algebra. It is very often stated that compactness of the group manifold is a lattice artifact because it could induce spurious effects in the approach to the continuum limit. In fact a global formulation of gauge theories also exists in the continuum in terms of loop variables [3]. Such a non-local description of gauge theories contains all their kinematical properties, though it is not obvious how to write down a classical action for loops. We shall consider the possibility that the global structure of gauge groups might play a role in the non-pertubative properties of gauge theories. But we must impose that any admissible lattice formulation does reproduce the results of the continuum non compact formulation in the pertubative regime.

There is a large degree of arbitrariness in the choice of action on the lattice. In particular the most general U(2) lattice action which can be built out of the elementary plaquette $P$ reads

$$S[U_P] = \sum_r \beta_r \, \text{Tr}_r \, U_P + \text{h.c.}$$
$$= \sum_{n,j} \beta_{n,j} \, e^{in\phi_P} \, \text{Tr}_j \, V_P + \text{h.c.} \quad (3)$$

where the sum runs over all irreducible representations (irreps) $r$ of U(2), which are the direct products of the U(1) irreps (labelled by an integer $n \in Z$) and SU(2) irreps (labelled by a non-negative integer or half-integer $j$), subject to the constraint $(-1)^{n+2j} = 1$, which is a consequence of Eq. (1).

It is generally believed that the choice of action is largely irrelevant once the gauge group is fixed, because it amounts to a modification of the regularization scheme which should not affect the continuum limit as long as weak coupling pertubation theory is applicable. In particular the U(2) lattice gauge theory is expected to reproduce the SU(2)×U(1) asymptotic scaling properties.

If we parametrize an SU(2) group element $V$ as a $2 \times 2$ unitary unimodular matrix, the Wilson action for the SU(2)×U(1) lattice gauge theory without matter fields reads:

$$S[U] = \sum_P b_1 \cos \phi_P + \frac{b_2}{4} \text{Tr} \, (V_P + V_P^\dagger) \quad (4)$$

Because of the decoupling between the U(1) and SU(2) gauge fields, its properties can be directly inferred from those of the U(1) and SU(2) lattice gauge theories with Wilson action, which are well-know [4, 5], and are summarized in Fig. 1. There is a U(1) first-order[1] critical line at $b_1^{cr} \sim 1.01$ and the renormalization group trajectories are the straight lines

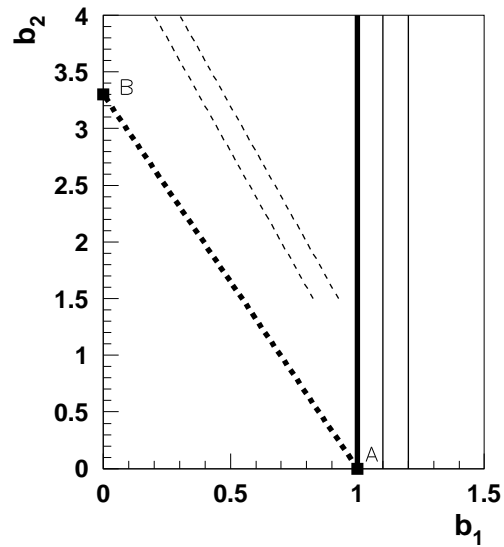

Figure 1: Critical curve and renormalization group trajectories of SU(2)×U(1) lattice gauge theory with Wilson action (solid lines) and expected results for U(2) lattice gauge theory with extended action (dashed lines).

parallel to the $b_2$ axis in the U(1) cold phase $b_1 > b_1^{cr}$.

They are labelled by a constant effective U(1) charge $e(b_1)$. In the compact formulation there is a natural dual description of massless free photons in terms of monopoles loops which explains the renormalization of the bare charge $b_1$ in the cold phase and the "confinement" transition through monopole condensation [11]. Along each renormalization group trajectory the physics is that of a confined SU(2) gauge theory which fixes the scale through the string tension. In the U(1) hot phase $b_1 < b_1^{cr}$ there is no continuum limit.

The simplest actions for the U(2) lattice gauge theory which have a naive continuum limit consistent with the two bare couplings of the SU(2)×U(1) gauge theory are obtained from Eq. (3) by retaining the first two terms consistent with the condition $(-1)^{n+2j} = 1$, namely $n = 2, j = 0$ and, $n = 0, j = 1$ or $n = 1, j = 1/2$:

---

[1] The transition is first-order at least for periodic boundary conditions, i.e. when the lattice is a torus $T^4$.

$$S[U] = \sum_P \quad b_1 \cos 2\phi_P$$
$$+ \frac{b_2}{3} \operatorname{Tr} V_P \operatorname{Tr} V_P^\dagger \quad (5)$$
$$S[U] = \sum_P \quad b_1 \cos 2\phi_P$$
$$+ \frac{b_2}{4} \operatorname{Tr} (e^{i\phi_P} V_P + e^{-i\phi_P} V_P^\dagger) \quad (6)$$

Eq. (5) is nothing but a lattice action for the SO(3)×U(1)/Z(2) gauge theory. The decoupling between the U(1) and SO(3) gauge fields on the lattice makes its asymptotic scaling properties in the continuum similar to SU(2)×U(1). However the SO(3) gauge theory does not exhibit confinement since there is a first-order critical point at $b_2^{cr} \sim 2.5$ [6] and the SO(3) lattice potential is screened at large distances. We have here a noteworthy example of the dependence of the non pertubative properties of a gauge theory upon the topological structure of its gauge group.

On the other hand action (6) introduces a coupling on the lattice between the U(1) and SU(2) gauge fields. It is usually expected that this coupling disappears in the continuum limit. The naive continuum limit of this action is, up to an additive constant:

$$S[U] = -a^4 \sum_{x,(\mu\nu)} \left(2b_1 + \frac{b_2}{2}\right) F_{\mu\nu}^2(x)$$
$$+ \frac{b_2}{4} \operatorname{Tr} W_{\mu\nu}^2(x) \quad (7)$$

where $a$ is the lattice spacing, $F_{\mu\nu}(x)$ and $W_{\mu\nu}(x)$ the U(1) and SU(2) field strength tensors respectively. In this limit the renormalized group trajectories in the $(b_1, b_2)$ plane are the straight lines with equations:

$$\frac{1}{g_1^2} = 2b_1 + \frac{b_2}{2} \quad (8)$$

They are sketched in Fig. 1. The analogy with SU(2)×U(1) would imply that these lines are labelled by an effective coupling $e(g_1)$ to a massless U(1) gauge field. Along these lines the physics should again be the same as for a confined SU(2) gauge theory.

The U(2) lattice gauge theory has been studied [7, 8] with the Wilson action, which corresponds to letting $b_1 = 0$ in Eq. (6). It is therefore not clear how to extract the continuum limit from these studies which were done in the early days of Monte-Carlo simulations of lattice gauge theories. However a convincing signal for a first-order transition was found for $b_2 \sim 3.3$ and some evidence was found for a decoupling between the U(1) and SU(2) gauge fields. This decoupling had also been established for the Wilson action of U(N) lattice gauge theory in the large N limit [9]. It was clearly recognized [7] that a determinant piece should be added to the Wilson action of U(2) lattice gauge theory[2] but the computing power was not available at the time. To our best knowledge this problem has not been addressed since then.

Our purpose is to report on a study of U(2) lattice gauge theory with extended action (6). In the next section we make a prerequisite step which is the determination of the phase diagram of such a lattice model. In this work we describe only the phase diagram in the first quadrant $(b_1 > 0, b_2 > 0)$ of the parameter plane. We find a first-order critical line which joins, as anticipated in Fig. 1, the points $A(1.0, 0.)$ and $B(0., 3.3)$. We then confirm [7, 8, 9] that the determinant of the Wilson loops can serve to characterize the transition. The determinant probes the central U(1) subgroup of U(2) and we can extract its lattice potential. We find that one phase is confined with a non-zero string tension and the other phase is deconfined with an effective U(1) charge. We determine the contour levels of these effective charges and find that, asymptotically, they are straigth lines, in agreement with the naive continuum analysis. We also study the lattice potential in the fundamental representation of U(2). In the deconfined U(1) phase we find a non-zero string tension which it is natural to associate to the SU(2) part of U(2). We study the renormalization flow of the bare coupling of the SU(2) subgroup by holding fixed the string tension and find that it is consistent with the contour levels of the effective U(1) charge and with universality.

In our conclusion we make some remarks about the limitations of this work and further possible extensions. Let us mention here that the phase diagram has also a rich structure in the quadrant $(b_1 < 0, b_2 > 0)$. The well-known hysteresis phenomenon at first-order transitions makes difficult a reasonably precise determination of this phase diagram. It remains to unravel whether the additional features are relevant to the continuum limit of the U(2) lattice gauge theory and how the phase structure is related to the monopole content of the model [12].

---
[2] Samuel has also made an analytic study of the large N limit of U(N) lattice gauge theory with fundamental and ajoint action terms [10], which corresponds to retaining the terms $n = 1$, $j = 1/2$ and $n = 0$, $j = 1$ in Eq. (3).

## 2 Phase diagram

In order to determine the phase diagram of the $U(2)$ lattice gauge theory with action (6) a standard Monte-Carlo simulation has been done on an $8^4$ lattice with periodic boundary conditions using a local heatbath algorithm. There is, perhaps, one technical point worth mentioning in the implementation of the heatbath algorithm. Because action (6) contains both determinant and trace terms, it proves more convenient to use a parametrization of the U(2) group which incorporates its quotient structure:

$$U = \begin{pmatrix} u_1 + i\, u_2 & u_3 + i\, u_4 \\ e^{i\theta}(-u_3 + i\, u_4) & e^{i\theta}(u_1 - i\, u_2) \end{pmatrix} \quad (9)$$

with $\quad u_1^2 + u_2^2 + u_3^2 + u_4^2 = 1$ and $0 \leq \theta < 2\pi$.

Indeed with this parametrization action (6) depends linearly upon $\cos\theta$, $\sin\theta$ and the $u_i$ variables. The U(2) Haar measure is still a direct product of the form $dU \sim d\theta d^4 u \delta(u^2 - 1)$. Therefore we can generate a U(2) group element by using standard or optimized heatbath algorithms for U(1) and SU(2).

This study has been restricted to the first quadrant ($b_1 > 0$, $b_2 > 0$) in the parameter plane. Data have been taken for several hundreds of points with 2000 iterations for each point after a cold start. Each run has been analyzed by the histogram technique with the PAW package [22]. In order to determine the phase diagram we have included in the analysis the measurements of the trace and determinant of the elementary plaquette. Average values and standard deviations are obtained by a gaussian fit to the histograms. Unthermalized data are spotted by a bad $\chi^2$ of the fit. Typically a few hundred of iterations are discarded for thermalization. The number is adjusted from the $\chi^2$. First-order transitions are located by looking for histograms with a two peak structure fitted by two gaussians.

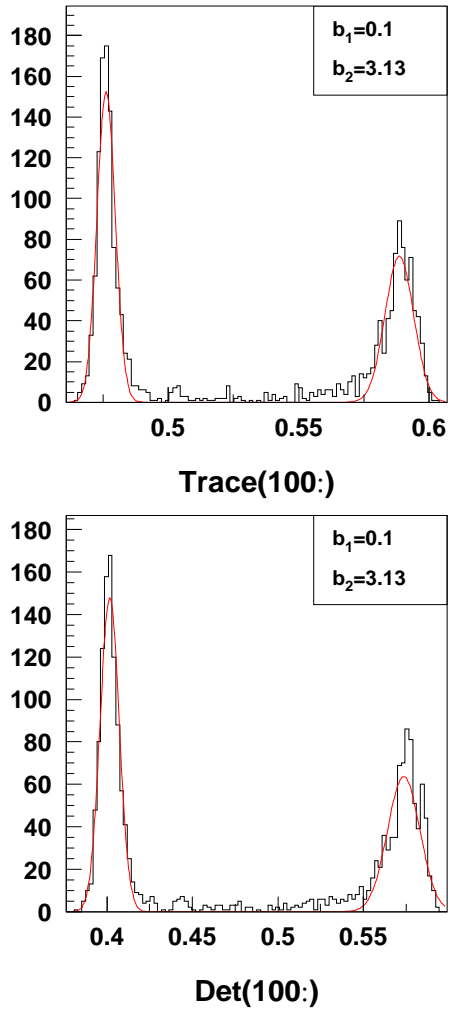

Figure 2: Typical histograms in the critical region.

Typical two-peak histograms and their fit are shown in Fig. 2 for the trace and determinant.

We have a very clear double-peak signal for a first-order transition in the $(b_1, b_2)$ plane. Critical couplings are defined by fits with two gaussians of equal weight. The critical couplings coincide for the trace and the determinant. Their values are displayed in Table 1. The first-order critical curve is shown in Fig. 3. It joins the U(1) critical point at $(1.0, 0.)$ to the U(2) critical point at $(0., 3.3)$ [7].

A by-product of analyzing first-order transitions with histograms is the direct measurement of the discontinuities of observables at the transitions. The gaps in the plaquette en-ergies can be read from Table 1 which contains the values of the jumps in the trace and determinant of the plaquette along the critical curves in the cold and hot phases. These gaps are also plotted in Fig. 4. We see clearly two regimes in the gaps if we follow the critical curve labelled by the coupling $b_1$. Starting from the U(1) critical point at $b_1 = 1.0$ the gap in the determinant stays rather constant down to $b_1 \sim 0.7$ where it departs from the U(1) value.

| $b_1$ | $b_2^c$ | $b_2^h$ | $\Delta T$ | $\Delta D$ |
|---|---|---|---|---|
| 0.0 | 3.28 | 3.35 | 0.117 | 0.180 |
| 0.1 | 3.14 | 3.20 | 0.117 | 0.180 |
| 0.2 | 2.99 | 3.04 | 0.112 | 0.174 |
| 0.3 | 2.84 | 2.88 | 0.106 | 0.165 |
| 0.4 | 2.69 | 2.72 | 0.084 | 0.138 |
| 0.5 | 2.53 | 2.55 | 0.057 | 0.102 |
| 0.6 | 2.35 | 2.36 | 0.035 | 0.072 |
| 0.74 | 2.0 | | 0.018 | 0.055 |
| 0.86 | 1.5 | | 0.010 | 0.052 |
| 0.94 | 1.0 | | 0.006 | 0.048 |
| 0.99 | 0.5 | | 0.003 | 0.045 |
| 1.01 | 0.0 | | 0.0 | 0.045 |

Table 1: Critical couplings of U(2) $8^4$ lattice gauge theory after a cold (c) and hot (h) start. $\Delta T$ and $\Delta D$ denote the jumps in the trace and determinant of the plaquette across the critical line.

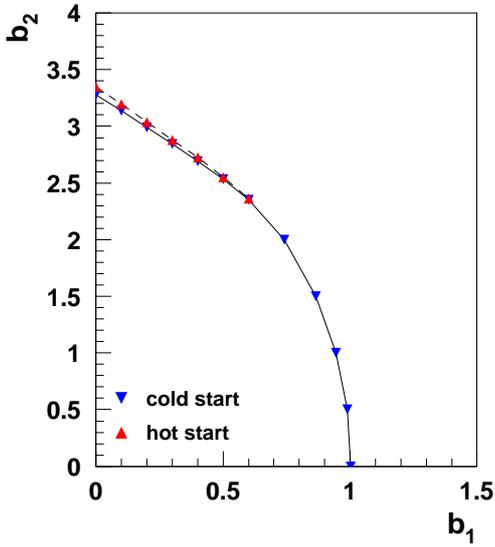

Figure 3: First-order critical curve of U(2) $8^4$ lattice gauge theory. Solid line corresponds to cold start, dashed line to hot start. Lines interpolate linearly between data points denoted by markers.

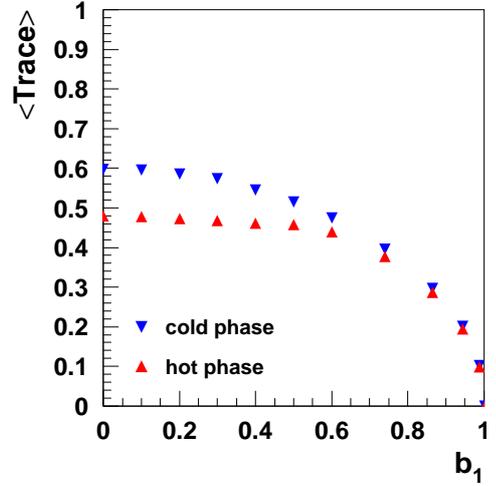

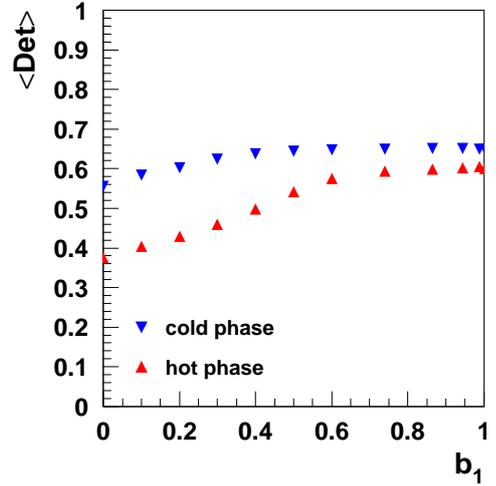

Figure 4: Gaps in the plaquette energy along the first-order critical curve for U(2) $8^4$ lattice gauge theory.

There is a well-known problem that we must worry about when studying first-order transitions which is the hysteresis phenomenon. The autocorrelation time for local Monte-Carlo algorithms is very long at a first order transition because any local algorithm requires the system to pass through the minimum separating the two peaks. This tunneling time increases exponentially with the gap between the peaks. The aforementionned widening of the gap makes the critical slowing down even worse for U(2) than for U(1). In order to assess the problem more quantitatively we have repeated the Monte-Carlo runs with a hot start for all points in the critical region. Everything else is unchanged: local heatbath algorithm, 2000 iterations, histogram analysis. The corresponding critical couplings are shown in Table 1 and plotted in Fig. 3. We observe that the hysteresis $\Delta b = b_{cr}^h - b_{cr}^c$ correlates well with the width of the gaps. A typical hysteresis cycle is shown in Fig. 5.

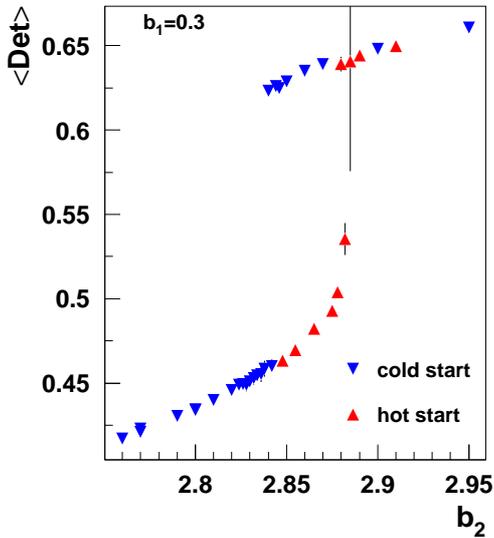

Figure 5: Typical hysteresis cycle in the critical region.

These hysteresis effects make a precise determination of the critical curve practically impossible with a local algorithm. The standard prescription to locate the phase transitions is to make runs from mixed starts which corresponds roughly to take the average of the two values listed in Table 1. The hysteresis induces systematic errors in the values of the gaps of Table 1 which are much larger than the statistical errors. We estimate these systematic errors to be of the order of 0.01. In the sequel our focus will be rather on the bulk properties of the phases which do not depend upon the precise location of the critical curve.

## 3    Order parameter

The critical branch found in the previous section goes through the U(1) critical point at $b_1 = 1.0$, $b_2 = 0$. Since only the determinant term in the action plays a dynamical role in the vicinity of this point, it is natural to expect that the order parameter of the phase transition of the U(1) subgroup probed by the determinant is also the order parameter of the U(2) phase transition along this branch. The order parameter of the U(1) phase transition is the string tension [4] which is zero in the cold phase and non-zero in the hot phase. Therefore we need to extract the string tension produced by the central U(1) subgroup of U(2) from the determinant of Wilson loops.

We use a method [13] which is very effective on intermediate size lattices and would be quite useful also on large lattices. In Monte-Carlo measurements of the string tension, one needs operators on the lattice which are usually built either from planar rectangular loops, or from thermal lines[4]. In both cases an extrapolation in one, say the time, direction is necessary to make contact with the physical observables. This extrapolation introduces an important source of systematic errors. There is a convenient way to circumvent this problem. It consists of building lattice operators which contain only one length scale. This criterion will be met if one uses loops inscribed on an hypercube. It is easy to form, with one-scale non-planar loops, ratios from which ultraviolet divergences can be removed by a charge renormalization only:

$$D_C(L, \beta) = \frac{< \det W(C, L) >}{< \det W(L, L) >^{P/4}} \qquad (10)$$

where $\beta$ denotes, collectively, the coupling parameters on the lattice, $\det W(C, L)$ is the determinant of the U(2) phase factor along the loop $C$ inscribed on an hypercube of side $La$ and $\det W(L, L)$ is the determinant along the square loop of side $La$ to the power $P/4$ where $PL$ is the perimeter of $C$ in lattice spacing units $a$. Then all perimeter and corner singularities are removed [15]. Wilson's area law gives the leading contribution to the ratio $D_C(L,\beta)$ and we can write, from dimensional analysis alone:

$$\begin{aligned} D_C(L,\beta) &= \exp\{-K_1\left(S_C - P/4\right)L^2 a^2 \\ &\quad - B_C + \cdots\} \end{aligned} \quad (11)$$

where $K_1$ is the string tension in the determinant representation, $S_C$ is the area of the minimal surface enclosed by the loop $C$ inscribed on the unit hypercube, $B_C$ is a constant related to the effective charge of the Coulomb part of the potential and the discarded terms decrease as $1/L^2$. The constants $B_C$ depend upon the intrinsic geometry of the loop $C$ but not[5] upon the scale $L$.

---
[4]For a recent review about string tension see [14] and references about standard algorithms therein.

| Loop | Description | $S(C)$ | $\Gamma(C)$ |
|------|-------------|--------|-------------|
| C1 | 1, 2, 3,-1,-2,-3 | 1.930 | 0.394 |
| C2 | 1, 2, 3, 4,-1,-2,-3,-4 | 3.014 | 0.714 |
| C3 | 1, 2, 3, 4,-2,-1,-4,-3 | 2.899 | 0.525 |

Table 2: Description of the set of non-planar loops included in the measurements. $\pm\mu$ means a displacement of $\pm 1$ in direction $\hat{\mu}$. $S(C)$ is the area of the minimal surface enclosed by loop $C$ and $\Gamma(C)$ a geometrical factor described in the text.

We have measured the ratios (10) for the set of three loops shown in Table 2. Hundred measurements have been done every other ten of the last thousand iterations of each run of the Monte-Carlo simulation described in the previous section. Since the runs are done on an $8^4$ lattice, finite-size effects are negligible only for measurements at scales $L = 1$ and $L = 2$. This is enough to get a sensible two-parameter fit of the constants $K_1 a^2$ and $B_C$ for each loop $C$ separately. Consistency can be judged by comparing the values of $K_1 a^2$ between all loops. We find a clear signal that the order parameter of the U(2) phase transition is indeed the string tension produced by the central U(1) subgroup of U(2). The string tension $K_1$ jumps discontinuously from zero to a non-zero value across the critical curve. A typical behavior of this string tension in the critical region is shown in Fig. 6. The value of $K_1 a^2$ is zero within statistical errors everywhere in the cold phase and for all three loops.

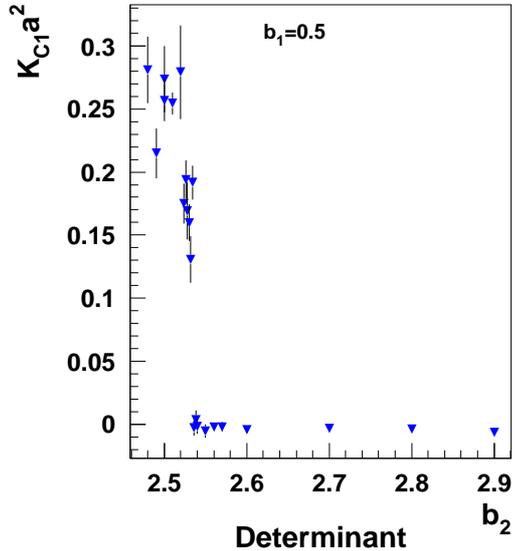

Figure 6: String tension $K_1$ of the central U(1) subgroup of U(2) in the critical region from the loop $C1$.

## 4 U(1) effective charge

Our introductory analysis of the relation between the U(2) lattice gauge theory and SU(2)×U(1) suggests that the continuum limit should be reached along the lines of constant U(1) effective charge in the cold phase. Therefore we have looked in this phase for the contour levels of the fitted parameters $B_C$ since they should describe the lattice potential in the determinant representation. However we must resort to some model if we want to label the contour levels with a loop-independent effective charge to a Coulomb field. A pertubative calculation has proved to fit the data quite well [13]. A first-order expansion gives for U(1)

$$B_C(\beta) = \frac{g_1^2}{4\pi}\Gamma(C) + O(g_1^4) \quad (12)$$

where $\Gamma(C)$ is a loop-dependent geometrical factor produced by the one-photon exchange between parallel edges of the loop. The expansion parameter $g_1$ is to be interpreted as an effective coupling which depends upon the coupling parameters $\beta$ on the lattice. The agreement between the fitted values of the effective charge $\alpha_1 = \frac{g_1^2}{4\pi}$ for different loops $C$ is a stringent test of the applicability of the pertubative approach. These contours are plotted in Fig. 7.

---
[5]We ignore possible logarithmic corrections which are very difficult to separate out anyhow.

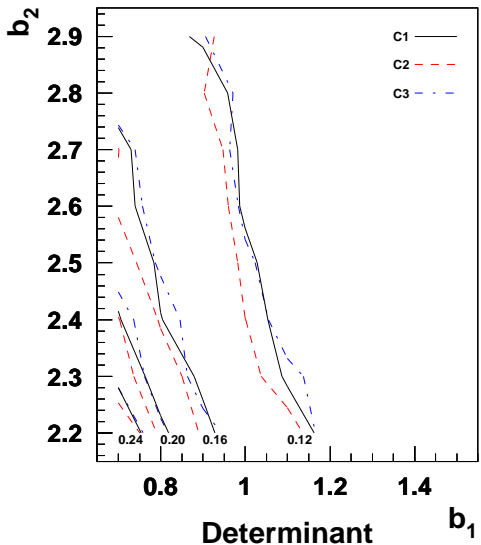

Figure 7: Contour levels $\alpha_1$=0.12, 0.16, 0.20 and 0.24 of the effective charge of the central U(1) subgroup of U(2) in the cold phase determined from the loops C1 (solid lines), C2 (dashed lines) and C3 (dash-dotted lines).

We have taken data in the cold phase on a coarse rectangular grid $0.7 \leq b_1 \leq 1.5$, $2.2 \leq b_2 \leq 2.9$ with spacing $\Delta b_1 = \Delta b_2 = 0.1$. Nethertheless we can see that the contour levels are qualitatively the same for all loops $C$ even if the roughness of the curves in Fig. 7 reminds us that the parametrizations of Eqs. (11,12) are only approximate.

It is also possible to make a model-independent analysis of the data and to study directly the contour levels of the ratios $D_C(L, \beta)$ or, equivalently, $V_C(L, \beta) = -\ln D_C(L, \beta)$. We refer loosely to $V_C$ as the lattice potential though it is not to be confused with the static potential. In the determinant representation, since the string tension $K_1$ is zero, we have $V_C(L) \approx B_C$. We exhibit in Fig. 8 some contour levels of $V_C$ extracted from the potential data at scale 1 where the signal-to-noise ratio is of course much higher.

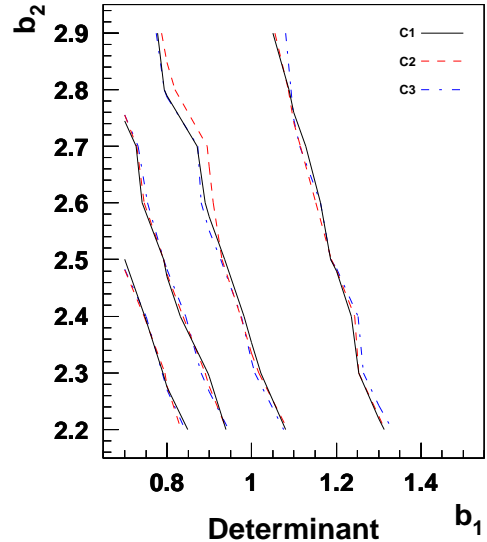

Figure 8: Contour levels $V_{C1}$=0.04, 0.05, 0.06 and 0.07 (solid lines), $V_{C2}$=0.068, 0.086, 0.102 and 0.120 (dashed lines), $V_{C3}$=0.053, 0.068, 0.080 and 0.095 (dash-dotted lines).

We observe that the contours coincide pretty well for all loops, once the level values are suitably chosen, and that they are close to parallel straight lines, in agreement with the naive continuum limit. A closer look shows that the parallelism is only approximate and that the slope of the contour levels tends to decrease when the potential decreases. For the lowest level the slope is about -2.7 which is still very different from the naive continuum limit. But reaching this limit requires taking $b_1$ and $b_2$ to $\infty$ and the pertubative corrections are known to be large on the lattice.

We have much more data in the critical region and we can attempt to map the contour levels in the vicinity of the critical curve inside the cold phase. A typical behavior of the effective charge $\alpha_1$ across the critical region is shown in Fig. 9. It can be seen that a linear fit can describe the data near the critical point.

The values of $\alpha_1$ along the critical curve inside the cold phase are listed in Table 3 for all loops we have studied. These values are within three standard deviations of their average $\alpha_1^c = 0.34$. The overall consistency among the loops and along the critical curve is pretty good and shows that the pertubative parametrization is at least a convenient way of describing the data. However we must keep in mind that the systematic errors due to the hysteresis phenomenon and the pertubative parametrization are much larger than the statistical errors listed in Table 3.

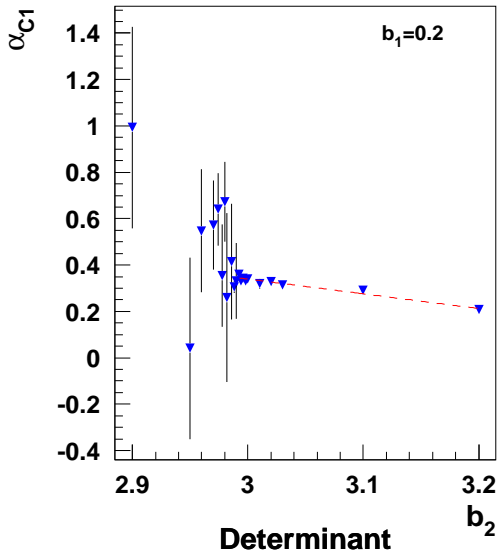

Figure 9: U(1) effective charge $\alpha_1$ in the critical region from the loop C1 and linear fit in the cold phase.

The same analysis applied to the potential data at scale 1 in the critical region yields completely similar results. We shall summarize these results by stating that our findings are consistent with the fact that the first-order critical curve is a contour level of the central U(1) effective charge and, more generally, a contour level of the ratios $D_C$ at scale 1.

Such limiting values might have a natural interpretation in the non-pertubative approach [17], based on the Nambu-Goto string model, which describes $B_C$ as a Coulomb-like correction induced by the quantum fluctuations of the string. This correction is universal in the sense that it does not depend upon the gauge group and upon the action of the effective string theory which describes the dynamics of the flux tube in gauge theories. But it does depend upon the dimensionality of space-time and the geometry of the loop $C$. The parameters $B_C$ have been calculated for arbitrary rectangular loops [18] but not for non-planar loops.

| $b_2$ | $b_1$ | Loop | $\alpha_1^c$ | $\pm$ |
|---|---|---|---|---|
| 3.28 | 0.00 | C1 | 0.377 | 0.011 |
|  |  | C2 | 0.361 | 0.010 |
|  |  | C3 | 0.361 | 0.014 |
| 3.14 | 0.10 | C1 | 0.344 | 0.010 |
|  |  | C2 | 0.345 | 0.007 |
|  |  | C3 | 0.359 | 0.008 |
| 2.99 | 0.20 | C1 | 0.340 | 0.007 |
|  |  | C2 | 0.322 | 0.005 |
|  |  | C3 | 0.351 | 0.010 |
| 2.84 | 0.30 | C1 | 0.323 | 0.007 |
|  |  | C2 | 0.312 | 0.007 |
|  |  | C3 | 0.332 | 0.008 |
| 2.69 | 0.40 | C1 | 0.326 | 0.010 |
|  |  | C2 | 0.315 | 0.008 |
|  |  | C3 | 0.331 | 0.010 |
| 2.54 | 0.50 | C1 | 0.331 | 0.012 |
|  |  | C2 | 0.324 | 0.012 |
|  |  | C3 | 0.337 | 0.013 |
| 2.35 | 0.60 | C1 | 0.332 | 0.010 |
|  |  | C2 | 0.313 | 0.008 |
|  |  | C3 | 0.333 | 0.011 |
| 2.00 | 0.74 | C1 | 0.331 | 0.009 |
|  |  | C2 | 0.320 | 0.008 |
|  |  | C3 | 0.344 | 0.010 |
| 1.50 | 0.86 | C1 | 0.341 | 0.011 |
|  |  | C2 | 0.338 | 0.008 |
|  |  | C3 | 0.362 | 0.010 |
| 1.00 | 0.94 | C1 | 0.329 | 0.010 |
|  |  | C2 | 0.333 | 0.008 |
|  |  | C3 | 0.333 | 0.009 |
| 0.50 | 0.99 | C1 | 0.345 | 0.011 |
|  |  | C2 | 0.337 | 0.009 |
|  |  | C3 | 0.357 | 0.012 |

Table 3: Values of the U(1) effective charge $\alpha_1$ along the critical curve from all loops $C$.

All the results of the last two sections have been obtained from measurements at scale 1 and 2 in lattice spacing units. A noteworthy consequence is the smallness of statistical errors and a very good signal. One might wonder how it is possible to extract physical quantities from so small lattice distances. But when evaluating the results with respect to conventional methods, one should take into account the absence of extrapolations, which is the heart of our technique.

## 5 Renormalization flow

We can repeat the analysis of the previous sections for the one-scale ratios built with the trace of Wilson loops in the fundamental representation of $U(2)$:

$$T_C(L,\beta) = \frac{<\operatorname{Tr} W(C,L)>}{<\operatorname{Tr} W(L,L)>^{P/4}} \qquad (13)$$

Again we write the same dimensional parametrization:

$$\begin{aligned}T_C(L,\beta) &= \exp\{-K_2(S_C - P/4)L^2 a^2 \\ &\quad - B'_C + \cdots\}\end{aligned} \qquad (14)$$

We have measured the ratios (13) for the same set of non-planar loops at the same scales $L = 1$ and $L = 2$ and fitted the constants $K_2 a^2$ and $B'_C$. Again we find that the string tension $K_2$ jumps discontinuously across the critical curve. But now its value stays different from zero in both phases. It is natural to associate the non-zero string tension in the deconfined $U(1)$ phase to the $SU(2)$ factor in $U(2)$. Fig. 10 shows two typical behaviors of $K_2 a^2$ in the critical region.

Along the line $b_1 = 0$, inside the cold phase, the value of $K_2 a^2$ is very small, $K_2 a^2 \leq 0.015$, which means that the correlation length $\xi = 1/\sqrt{K_2}$ is larger than the lattice, $\xi \geq 8a$. This finite-size effect explains the inconclusiveness of early studies with Wilson action [7] about a non-abelian string tension. For $b_1 = 0.4$, near the critical point, we have $K_2 a^2 \sim 0.09$ and a clear signal for a non-abelian string tension. However the value of $K_2 a^2$ decreases rapidly along the line $b_1 = 0.4$ and at $b_2 = 2.8$ the correlation length $\xi$ is again larger than the lattice. Therefore a required step, before attempting the study of the continuum limit of the $U(2)$ lattice gauge theory, is to determine the scaling window of our $8^4$ lattice. The most direct method is to use the contour levels of $K_2 a^2$. We define the scaling window, somewhat arbitrarily, by demanding that the correlation length $\xi$ be smaller than half the lattice, $K_2 a^2 \geq 0.06$, to avoid finite size effects. Some contour levels are displayed in Fig. 11.

The same observations that we made about the rough appearance of the contour levels of the $U(1)$ effective charge apply to Fig. 11. We expect the contours to become parallel to the $b_1$ axis for large $b_1$ where one should recover the values of the $SU(2)$ string tension with

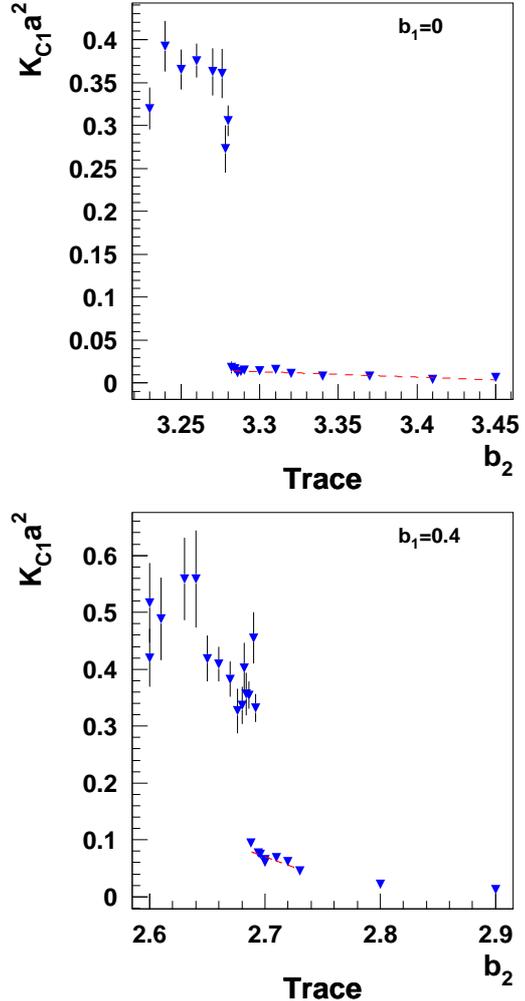

Figure 10: String tension $K_2$ in the critical region from the loop C1 and linear fit in the cold phase.

Wilson action. Indeed, in the limit $\frac{b_1}{b_2} \to \infty$, the $U(1)$ variables $\phi$ are driven up to a gauge transformation into the $Z(2)$ subgroup which, by the topological structure of the group $U(2)$, is also the center of $SU(2)$. Therefore, in this limit, the variables $\phi$ can be absorbed in the invariant $SU(2)$ measure and the $U(2)$ model with action (6) goes over into the usual $SU(2)$ theory with Wilson action. The levels have not yet entered this asymptotic regime at $b_1 = 1.5$ but they are consistent with the $SU(2)$ data.

On the other hand, towards smaller $b_1$, the contour levels hit the critical curve. We note from Fig. 10 that a linear fit can describe the data near the critical point in the cold phase. The values of $K_2 a^2$ along the critical curve inside the cold phase are listed in Table 4 for each loop $C$.

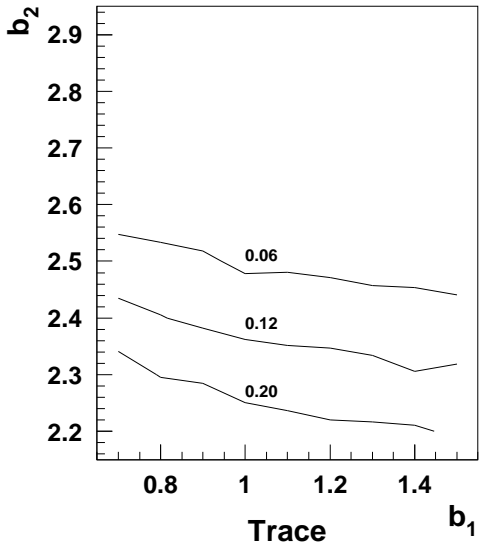

Figure 11: Contour levels $K_2 a^2 = 0.06$, 0.12 and 0.20 in the cold phase determined from the loop C1.

| $b_2$ | $b_1$ | Loop | $K_2^c a^2$ | $\pm$ |
|---|---|---|---|---|
| 3.28 | 0.00 | C1 | 0.014 | 0.002 |
|  |  | C2 | 0.012 | 0.002 |
|  |  | C3 | 0.021 | 0.003 |
| 3.14 | 0.10 | C1 | 0.016 | 0.002 |
|  |  | C2 | 0.014 | 0.002 |
|  |  | C3 | 0.023 | 0.003 |
| 2.99 | 0.20 | C1 | 0.024 | 0.002 |
|  |  | C2 | 0.022 | 0.002 |
|  |  | C3 | 0.036 | 0.003 |
| 2.84 | 0.30 | C1 | 0.038 | 0.003 |
|  |  | C2 | 0.041 | 0.003 |
|  |  | C3 | 0.062 | 0.004 |
| 2.69 | 0.40 | C1 | 0.087 | 0.005 |
|  |  | C2 | 0.095 | 0.005 |
|  |  | C3 | 0.134 | 0.007 |
| 2.54 | 0.50 | C1 | 0.143 | 0.006 |
|  |  | C2 | 0.153 | 0.006 |
|  |  | C3 | 0.236 | 0.013 |
| 2.35 | 0.60 | C1 | 0.238 | 0.008 |
|  |  | C2 | 0.281 | 0.035 |
|  |  | C3 | 0.476 | 0.046 |
| 2.00 | 0.74 | C1 | 0.590 | 0.103 |

Table 4: Values of the string tension $K_2 a^2$ along the critical curve from each loop $C$.

The values $K_2^c a^2$ give the intersection points of the contour levels with the critical curve. In particular the scaling window of the $8^4$ lattice begins, along the critical curve, around $b_2 \sim 2.8$. We get a good signal-to-noise ratio only down to $b_2 \sim 2.3$ where we enter the strong coupling regime.

We are now in a position to test whether the renormalization flow of the U(2) lattice gauge theory with action (6) is consistent with the lines of constant U(1) effective charge. We may do it with a reasonable accuracy only in the vicinity of the critical curve where we can use the values in Tables 3 and 4. Fig. 12 displays the values of $K_2^c a^2$ along the critical curve on a logarithmic scale as a function of $b_2 = \frac{4}{g_0^2}$. Here $g_0$ is the bare coupling constant of the SU(2) gauge field in the continuum limit.

If confinement persists in the continuum limit, the weak coupling behavior of $K_2 a^2$ should follow the prediction of the renormal-

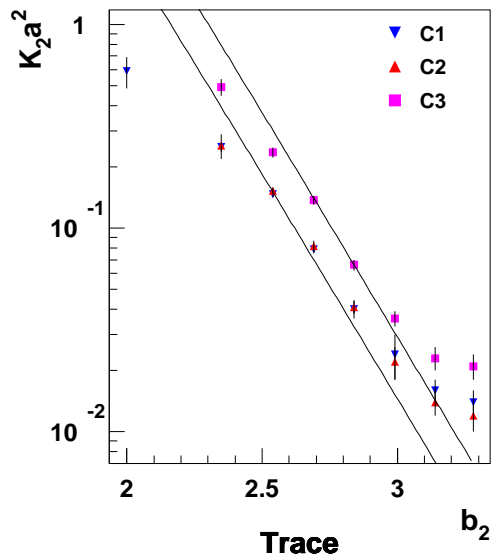

Figure 12: String tension $K_2$ in the cold phase along the critical curve.

ization group:

$$K_2 a^2 = \frac{K_2}{\Lambda^2} \exp\left[-\frac{1}{\beta_0 g_0^2}\right] \left(\beta_0 g_0^2\right)^{\frac{-\beta_1}{\beta_0^2}} \quad (15)$$

where $\beta_0$ and $\beta_1$ are the usual first two coefficients of the $\beta$ function for the SU(2) gauge theory and the renormalization scale $\Lambda$ depends upon the ratio $t = \frac{b_2}{b_1}$. $\Lambda(0)$ is the renormaliztion scale of the SU(2) lattice gauge theory with Wilson action and the dependence $\Lambda(t)/\Lambda(0)$ is in principle calculable in pertubation theory. The prediction of Eq. (15) is shown in Fig. 12 as the standard band representing values of the parameter $\Lambda$ in the range

$$\frac{\sqrt{K_2}}{\Lambda} = 145 \pm 25 \quad (16)$$

We observe that the values $K_2^c a^2$ extracted from the loops C1 and C2 agree with their average within two standard deviations but there is a significant discrepancy with the values extracted from the loop C3. The sizeable deviation of the loop C3, which explains the rather large (estimated) error in Eq. (16), is a hint that the systematic errors induced by the hysteresis are much larger than the statistical errors listed in Table 4. Nethertheless the string tensions extracted from all three loops have the same scaling behavior and there is an overall qualitative agreement of our data with asymtotic scaling in the expected window. Achieving quantitative agreement will require not only determining a contour level very accurately outside the critical region but also either very large lattices or some redefinition of the coupling constant $g_0$ [19].

## 6 Universality

Usually testing universality of a lattice gauge model with several parameters requires measuring a few asymptotic quantities such as a string tension $K$ or a glueball mass $M$ and checking that their ratios stay constant inside the scaling window of the model. Such observables are defined in terms of long-distance correlations, even in the continuum, and are very difficult to extract on the lattice. The ratio method of Creutz [16] is a technique to study scaling and universality with short-distance operators on the lattice from which singularities can be removed by renormalization of the bare couplings only. The one-scale non-planar ratios are best suited for such studies.

Let $R_C(L, \beta)$ be such a ratio where $L$ is the scale in lattice spacing units $a$ and $\beta$ denotes all the coupling parameters of the model. Inside the scaling window of the model the ratio $R$ must satisfy the homogeneous renormalization group equations. Then a curve of constant ratio define some scale $\lambda$ in physical units of $\xi = K^{-1/2}$:

$$R_C(L, \beta(L, \lambda)) = c(\lambda) \quad (17)$$

Taking the derivative of $R$ with respect to $L$, at fixed $\lambda$ defines a Beta-function. In other words the renormalization prescription is to set the ratio $R$ at the physical scale $\lambda$ to a fixed value $c$ for all values of the cut-off $a$.

It follows that, in a lattice gauge model which has only one bare coupling in the continuum limit, the contour levels of the ratios $R_C(L, \beta)$ must coincide in the scaling domain for all loops $C$, all scales $L$ and all irreducible representations of the gauge group. If there are several bare couplings in the continuum limit, the contours may depend on what representation defines the ratios $R_C(L, \beta)$. We have already determined in section 4 the contour levels of the ratios $D_C(L, \beta)$ at scale 1 and checked that they do not depend on the loop $C$. Fig. 13 displays some contour levels of the lattice potential at scale 1 in the fundamental representation of U(2):

$$V_C(L, \beta) = -\ln T_C(L, \beta) \quad (18)$$

We observe again that the contours coincide for all loops $C$. Our statistics are not quite sufficient to check quantitatively that the contour levels stay the same for the ratios $T_C(L)$ at scale 2. But Fig. 11 shows at least a qualitative consistency with the contours of the string tension $K_2$ which fixes the scale $\lambda$.

The contour levels of the ratios $T_C(L, \beta)$ differ from the contours of the ratios $D_C(L, \beta)$. This difference is the signature of the decoupling of the SU(2) and U(1) gauge fields. We have shown that the contours in the determinant representation of U(2) characterize the U(1) subgroup. We can test whether the contour levels in the trace representation of U(2) characterize the SU(2) subgroup by measuring the one-scale non-planar ratios in the adjoint

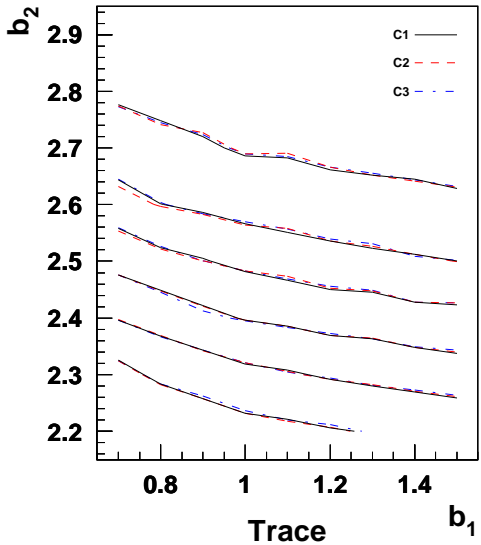
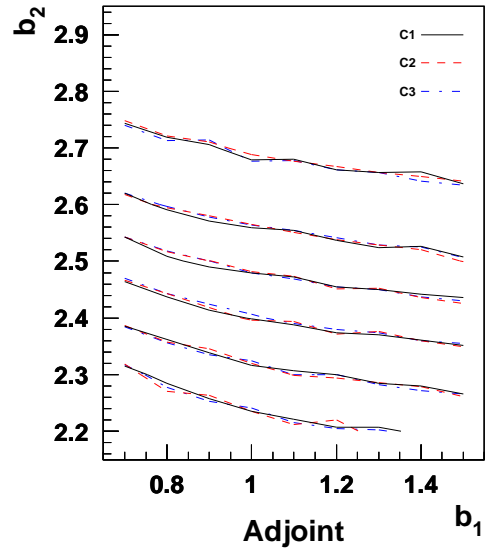

Figure 13: Contour levels $V_{C1}$=0.16, 0.19, 0.22, 0.26, 0.31 and 0.37 (solid lines), $V_{C2}$=0.290, 0.352, 0.413, 0.500, 0.610 and 0.750 (dashed lines), $V_{C3}$=0.232, 0.280, 0.330, 0.400, 0.485 and 0.590 (dash-dotted lines).

Figure 14: Contour levels $V_{C1}$=0.405, 0.475, 0.54, 0.625 0.73 and 0.845 (solid lines), $V_{C2}$=0.735, 0.88, 1.02, 1.195, 1.425 and 1.705 (dashed lines), $V_{C3}$=0.595, 0.705, 0.82, 0.955, 1.15 and 1.35 (dash-dotted lines).

representation of U(2):

$$A_C(L, \beta) = \frac{< \text{Tr}_A W(C, L) >}{< \text{Tr}_A W(L, L) >^{P/4}} \qquad (19)$$

Indeed these ratios depend only upon the SU(2) subgroup of U(2). If the continuum limit of the U(2) lattice gauge theory is the SU(2)×U(1) gauge theory, the adjoint contours must coincide with the trace contours in the scaling region. We exhibit in Fig. 14 some contour levels of the ajoint lattice potential at scale 1, $V_C(L, \beta) = -\ln A_C(L, \beta)$. The displayed level values have been chosen so as to make the contours in Fig. 14 coincide pretty well with the contours in Fig. 13.

## 7 Conclusion

All the results of our Monte-Carlo simulation provide evidence that the U(2) lattice gauge theory with action (6) does admit a continuum limit despite the existence of a first-order critical line in the phase diagram. Moreover all our data are consistent with the fact this continuum limit is the symmetric SU(2)×U(1) gauge theory. We have characterized the decoupling inside the cold phase of the central U(1) subgroup of U(2) by showing the existence of two distinct families of contour levels. We have shown that the curves of the first family are labelled by a dimensionless effective charge which can be interpreted as measuring the coupling to a free massless U(1) gauge field in the continuum. We have also shown that the curves of the second family are labelled by a scale measured in units of an SU(2) string tension. We have verified universality in the pertubative regime. What we have not checked completely is universality in the non-pertubative regime. This would require more statistics for a better determination of the U(1) and SU(2) contour levels at larger scales and larger lattices for calculating the spectrum. Even if the result is highly probable this test is in principle needed to prove that the continuum limit of the U(2) lattice gauge theory is indeed the symmetric SU(2)×U(1) gauge theory.

We have restricted this study to the upper right quadrant ($b_1 \geq 0, b_2 \geq 0$) of the parameter plane because we expected, from our preliminary analysis, to find the SU(2)×U(1) continuum limit in this domain. However the structure of the phase diagram of the model is much more complex and extends outside the first quadrant. The first-order critical curve continues beyond the point $B(0.0, 3.3)$ in the quadrant $(b_1 < 0, b_2 > 0)$ where hysteresis phenomena still worsen and obscure the interpretation of the data. Maybe using multicanonical algorithms [20] will improve the situation. Moreover the existence of another critical branch can be deduced as follows [10].

Let us consider the partition function of the U(2) lattice model with action (6):

$$Z(b_1, b_2) = \int D[U] \exp \sum_P b_1 \det U_P + \frac{b_2}{4} \operatorname{Tr} U_P + h.c. \quad (20)$$

We can choose, on a lattice with free boundary conditions, a set of links $\mathcal{L}$ such that each plaquette of the lattice contains exactly one link in $\mathcal{L}$. The integration measure is invariant with respect to left or right multiplication of any link by a U(2) group element. It follows in particular that the partition function is invariant under the change of variable $U_l \to e^{i\phi} U_l$ for all links $l$ in $\mathcal{L}$, which implies the identity:

$$Z(b_1, b_2) = Z(b_1 e^{2i\phi}, b_2 e^{i\phi}) \quad (21)$$

Letting $\phi = \pi$, the identity (21) implies that the phase diagram is symmetric with respect to the $b_1$ axis and we can restrict the study to $b_2 \geq 0$. Letting $\phi = \frac{\pi}{2}$ we have also

$$Z(b_1, 0) = Z(-b_1, 0) \quad (22)$$

which implies the existence of a critical point $A'$ at $(b_1, b_2) = (-1.0, 0.0)$ with the same properties as the U(1) critical point $A(1.0, 0.0)$. There must be another first-order critical branch which springs from $A'$, with the same slope as the first branch at $A$, and continues towards smaller negative $b_1$.

We have checked that this second critical branch does exist. We find that the gaps decrease along this branch from the U(1) value towards zero. Therefore the properties of the two branchs are markedly different and there is a possibility that the second branch may become second-order at some point. We hope to clarify in another report the nature and order parameter of this phase transition and, more generally, the properties of the phase diagram in the whole upper left quadrant.

The different global structures of the U(2) and SU(2)×U(1) gauge theories can be described in terms of monopoles [12]. There can be two types of monopoles in the U(2) gauge theory associated to the Z(2) intersection of the U(1) and SU(2) subgroups. There has been recently much effort [21] devoted to understanding the role of monopoles and boundary conditions in the phase transition of compact U(1) lattice gauge theory. Enlarging the study to U(2) lattice gauge theory would undoubtedly shed a new light on these questions.

# References

[1] L. Michel, *Rev. Mod. Phys.* **52**, 617 (1980) and references therein.

[2] K. Wilson, *Phys. Rev.* **D10**, 2445 (1974).

[3] S. Mandelstam, *Phys. Rev.* **175**, 1580 (1968); C.N. Yang, *Phys. Rev. Lett.* **33**, 445 (1974); A.M. Polyakov, *Phys. Lett.* **B82**, 247 (1979); Yu. M. Makeenko, A. A. Migdal, *Phys. Lett.* **B88**, 135 (1979).

[4] T.A. DeGrand and D. Toussaint, *Phys. Rev.* **D22**, 2478 (1980); J. Jersak, T. Neuhaus and P.M. Zerwas, *Phys. Lett* **B133**, 103 (1983).

[5] M. Creutz, *Phys. Rev. Lett.* **45**, 313 (1980); M. Creutz, L. Jacobs and C. Rebbi, *Physics Reports* **95**, 201 (1983) and references therein.

[6] G. Bhanot and M. Creutz, *Phys. Rev.* **D24**, 3212 (1981).

[7] M. Creutz and K.J.M. Moriarty, *Nucl. Phys.* **B210** [FS6], 59 (1982).

[8] J.M. Drouffe and K.J.M. Moriarty, *Z. Phys.* **C24**, 395 (1984).

[9] F. Green and S. Samuel, *Nucl. Phys.* **B194**, 107 (1982).

[10] S. Samuel, *Phys. Lett.* **B112**, 237 (1982).

[11] T. Banks, R. Myerson and J.B Kogut, *Nucl. Phys.* **B129**, 493 (1977).

[12] E. Corrigan and D. Olive, *Nucl. Phys.* **B110**, 237 (1976).

[13] D. Arnaudon and C. Roiesnel, *Phys. Lett.* **B187**, 153 (1987).

[14] E. Marinari, M.L. Paciello, B. Taglienti, Preprint ROME-1098-1995, submitted to Mod. Phys. Lett. A, e-Print Archive: hep-lat/9503027.

[15] V.S. Dotsenko and S.N. Vergeles, *Nucl. Phys.* **B169**, 527 (1980); R.A. Brandt, F. Neri and M. Sato, *Phys. Rev.* **D24**, 879 (1981).

[16] M. Creutz, *Phys. Rev.* **D23**, 1815 (1981).

[17] M. Lüscher, K Symanzik and P. Weisz, *Nucl. Phys.* **B173**, 365 (1980).

[18] K. Dietz and T. Filk, *Phys. Rev.* **D27**, 2944 (1983).

[19] G.P. Lepage and P.B. Mackenzie, *Phys. Rev.* **D48**, 2250 (1993).

[20] B.A. Berg and T. Neuhaus, *Phys. Lett.* **B267**, 249 (1991).

[21] Th. Lippert, G. Bhanot, K. Schilling and P. Ueberholz, *Nucl. Phys.* **B30** (Proc. Suppl.), 912 (1993); C. B. Lang and T. Neuhaus, *Nucl. Phys.* **B431**, 119 (1994); M. Baig and H. Fort, *Phys. Lett.* **B332**, 428 (1994); W. Kerler, C. Rebbi and A. Weber, *Phys. Rev.* **D50**, 6984 (1994).

[22] R. Brun, O. Couet, C. Vandoni and P. Zanarini, *PAW users guide*, Program Library **Q120** and **Y251**, CERN, 1991.